\documentclass[journal,transmag]{IEEEtran}

\ifCLASSINFOpdf

\else

\fi
\usepackage{amssymb,amsmath,epsfig}
\makeatletter
\def\maketag@@@#1{\hbox{\m@th\normalfont\normalsize#1}}
\makeatother
\usepackage{graphicx}
\usepackage{epstopdf}
\usepackage{microtype}
\usepackage{cite}
\usepackage{url}
\usepackage[none]{hyphenat}
\usepackage{booktabs}
\usepackage[linesnumbered,ruled,vlined]{algorithm2e}
\usepackage{algorithmicx}
\usepackage{algpseudocode}
\usepackage{multirow}
\usepackage{nicefrac}
\everymath{\displaystyle}
\usepackage{color}
\usepackage[font=footnotesize]{caption}
\usepackage[linesnumbered,ruled,vlined]{algorithm2e}
\usepackage{algorithmicx}
\graphicspath{{Schematics/}} 
\usepackage{algpseudocode}
\usepackage{url}
\usepackage{hyperref}
\begin{document}

\title{Information Harvesting for Far-Field Wireless Power Transfer}

\author{\IEEEauthorblockN{Mehmet C. Ilter\IEEEauthorrefmark{1},
		Risto Wichman\IEEEauthorrefmark{1}, 
		Mikko Saily\IEEEauthorrefmark{2}} and Jyri H{\"a}m{\"a}l{\"a}inen\IEEEauthorrefmark{3}
	\IEEEauthorblockA{\IEEEauthorrefmark{1}Department of Signal Processing and Acoustics, Aalto University, 02150 Espoo, Finland}
	\IEEEauthorblockA{\IEEEauthorrefmark{2}Network and  Architecture  Group,  Nokia  Bell  Labs,  02610  Espoo,  Finland}
	\IEEEauthorblockA{\IEEEauthorrefmark{3}Department of Communications and Networking, Aalto University, 02150 Espoo, Finland}}%

\IEEEtitleabstractindextext{%
\begin{abstract}
Considering ubiquitous connectivity and advanced information processing capability, huge amount of  low-power IoT devices are deployed nowadays and the maintenance of those devices which includes firmware/software updates and recharging the units has become a bottleneck for IoT systems. For addressing limited battery constraints, wireless power transfer is a promising approach such that it does not require  any physical link between energy harvester and power transfer. Furthermore,  combining wireless power transfer with information transmission has become more appealing. In the systems that apply radio signals the wireless power transfer has become a  popular trend to harvest  RF-radiated energy from received information signal in IoT devices. For those systems,  design frameworks mainly deal with the trade-off between information capacity and energy harvesting efficiency. Therein various signaling design frameworks have been proposed for different system preferences between power and information. In addition to this, protecting the information part from potential eavesdropping activity in a service area introduces  security considerations for those systems.  In this paper, we propose a novel concept, Information Harvesting, for  wireless power transfer systems. It introduces a novel protocol design from opposite perspective compared to the existing studies. Particularly,  Information Harvesting aims to transmit  information  through existing wireless power transfer mechanism without interfering/interrupting power transfer.  To do so, only intended information receivers are able to decode the transmitted  information after they initiate an information transfer process at wireless power transmitter while energy harvesters continue energy harvesting procedure without noticing this procedure. Considering the diversity of IoT networks and the availability of wireless power transfer infrastructure, proposed Information Harvesting principles  may turn out to a pivotal methodology especially for the cases where  a large number of IoT devices require the  software/firmware updates along with periodical battery recharging needs.
\end{abstract}

}

\maketitle

\IEEEdisplaynontitleabstractindextext

\IEEEpeerreviewmaketitle

\section{Introduction}

\IEEEPARstart{T}he Internet of Things (IoT) creates the potential for new generation industry use cases  and smart cities. With uniquely identifiable devices capable of communicating through wireless environment, billions of devices are able to sense and interact with everything and everyone. The vision of IoT is set to be a communication platform for broadcast and point-to-point communication. From other aspect, wireless power transfer enables the next stage in the current consumer electronics
revolution, including battery-less sensors, passive RF identification (RFID), passive wireless
sensors, and machine-to-machine solutions. The  origin of wireless power transmitted by using RF spectrum was pioneered by Tesla and it is based on radiative energy transfer through a medium (i.e. air) with the help of antennas
\cite{Hui2014}. 

There are two types of power transfer: near-field power  and far-field power transfer. The first one depends on inductive/capacitive coupling between energy harvester and power transmitter so  its range is less than a meter. The latter one exploits radio frequency transmission  and due to propagation characteristics of RF signals,  it can be used over longer ranges \cite{CostAction2017}. For instance, prototype RF-based energy harvesting circuits operating below 1 GHz  are capable of harvesting microwatts to milliwatts of power over the range of 10 m with similar transmit power of a Wi-Fi router \cite{ng2019wireless}.

The main objective of radiative power transfer system is to maximize the harvested DC power subject to a given transmit power constraint, in other words,  enhancing the end-to-end power transfer efficiency. Designing an efficient rectenna, the unit that consists of a rectifier and an antenna, to maximize RF-to-DC conversion has been the traditional line of research in earlier power transfer literature \cite{Clerckx2019}. Furthermore, signal design for wireless power mechanism is getting more popular thanks to the availability of wider range of optimization techniques. For instance, different from grid-based and lattice-based constellations in conventional communication systems where lower peak-to-average power ratio is typically desired, real Gaussian signals, flash signaling and linear frequency-modulated signals  are the preferred constellations for energy harvesting mechanisms since higher PAPR values result in higher  DC voltage by a rectifier leading to better end-to-end power efficiency \cite{Clerckx2019} {at the cost of higher complexity and power consumption in the transmitter}.

Rather than radiating only energy into  a particular service area, combining RF-based
power transfer with simultaneous information transmission has become appealing with the recently introduced simultaneous wireless information and power transfer (SWIPT) concept. In SWIPT systems, the power and information components of the transmitting signal are separable by using the
energy domain (power splitting), the time domain (time splitting) and the space domain (antenna
splitting) \cite{Liu2016}. For those systems, the receiver architecture is generally optimized for increasing  total harvested energy while decoding information with minimum power consumption at the same time  without considerable loss in achievable data rates. Therefore, a trade-off exists between information transfer and energy transfer in such systems.

In practice, sensitivities of information receivers  and energy harvesters are quite different. Specifically, the minimum requirement for the received power at an information receiver is around –60 dBm, while that of an energy harvester is –10 dBm. {For instance under Friis assumption, 2.4 GHz frequency introduces a decay of 40 dB decay per meter.} Therefore, energy harvester  is usually located closer to wireless power transceiver than  information receivers. However, this setting may create a potential eavesdropping activity where receiver close by energy harvester can act as a malicious device who exploits better received signal statistics resulting in a high risk of eavesdropping and  information interception. Similar phenomenon is also  observed in power domain-non-orthogonal multiple access systems where successive interference cancellation techniques are implemented in the nearby users in order to eliminate other users' symbols thanks to having better channel conditions in the process of extracting its own intended signal from superposed transmitting symbol. Therefore, nearby users have an advantage to decode far user information as well and there is a need to combat against eavesdropping risk of nearby users by applying physical layer security schemes.

Physical layer (PHY)-security techniques which mainly turn the randomness in wireless channels into an advantage by exploiting  the physical aspects of communication channels between the nodes provide a powerful tool to achieve  secure communication. Although encryption mechanism is quite useful for protecting ongoing data transmission in many cases, considering the diversity in protocols and networks in 5G and beyond,  those mechanism might not be suitable for all available cases.  Even with limited computational capability, it was shown  that  the PHY-security is able to provide secure communications for a massive number of IoT devices \cite{Wang2019}.

Due to huge amount of devices, IoT networks require frequent software updates which include security updates, bug fixes, and software extensions. From this perspective, there is considerable amount of data transmission procedure occurring periodically. For those activities, encryption techniques can  contribute  up to 30\% of total energy consumption \cite{bauwens2020}.  Furthermore, most IoT devices/sensors are simple nodes having limited computational capabilities so the complexity of encryption-decryption procedures are beyond the capabilities of those devices. It is claimed that  no encryption is required in integrated/tiered IoT devices \cite{GSMA2014} and  SigFox protocol \cite{tournier2020survey}. Those trends can be easily verified by checking the statistics about existing  IoT traffic showing that the big portion of IoT traffic is unencrypted \cite{jeyanthi2019ubiquitous}. In addition, even encrypted data might not be enough if malicious users have enough time to analyze the traffic patterns to obtain the information. Therefore additional/complimentary solution is required.

To tackle these challenges this paper introduces Information Harvesting (IH) in wireless power transfer mechanism,  which   proposes data communication ability on top of existing  power transfer procedure; in other words,  wireless power transfer is partnered with a information transfer protocol which allows secure  transfer. The pivotal point of the proposed protocol is that the power receivers are not aware of information broadcasting and it is required to use a transmitter entity for the purpose of transmitting broadcasting packets without disturbing ongoing power transfer. From this perspective, IH can dissolve wireless power transfer and information broadcasting in one mechanism. This mechanism can be exploited through updating low complexity devices without any encryption (or low level of encryption for  necessary cases) by utilizing existing wireless power transceiver unit originally deployed for charging power receivers. 

{The rest of the paper is organized as follows. In Section II, the system components of IH mechanism are introduced. Then, the protocol details of  IH are represented in Section III. Then, Section IV introduces two different practical realizations, which are spatial modulation and analog beamforming, respectively. Besides,  some additional techniques which increase the physical layer security of the proposed protocol are also described therein. Section V represents the feasibility of spatial modulation based-information harvesting with simulation results generated for an indoor environment. Section VI concludes the paper.}

\section{System Components}
\begin{figure}[b]
\centering
\includegraphics[width=0.48\textwidth]{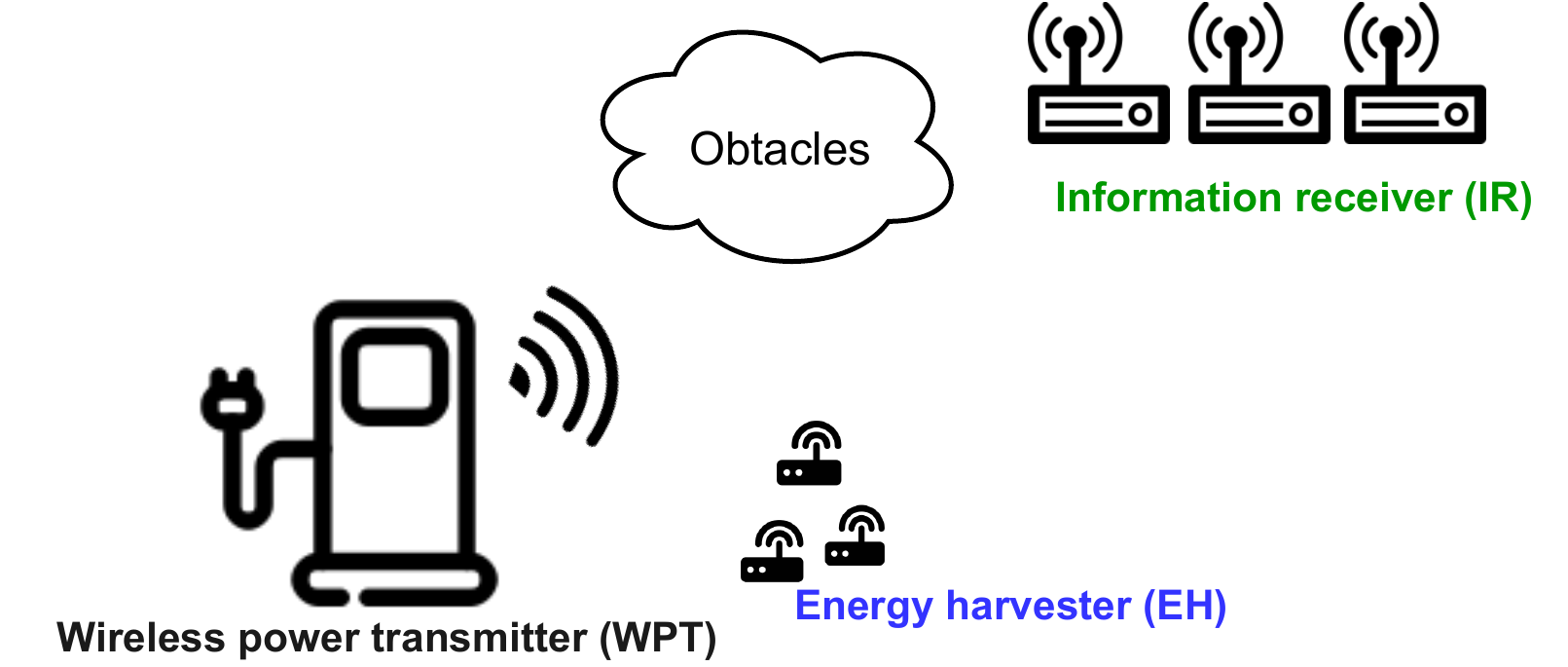}
\caption{The overall diagram of Information Harvesting system components.}
\label{fig:systeminformationHarvesting}
\end{figure}

The system components of IH are illustrated in Fig.~\ref{fig:systeminformationHarvesting}.  Herein, wireless power transceiver (WPT)  serves as a module supplying power transfer to a portable power receiver device(s), energy harvester (EH) and also providing information transfer by utilizing  information seeding to a portable information receiver device, information receiver (IR).  The IR contains receive antenna chain where single/multiple receive antenna(s) is(are) deployed and the energy harvester consists of a receive antenna chain, a battery charging unit and a battery. The key motivation herein is to design \textit{ information seeding} and \textit{information harvesting} cycles so that  the EHs are not able to detect the on-going information transfer and  in case of this activity detected by EHs, they are not able to decode the embedded  information. 

From this point of view, {information seeding} protocol lies on creating variations in transmitter entity in WPT with respect to information stream transmitted to IR. By doing so, broadcasting channels between WPT and IR can be configured by exploiting existing wireless power transfer mechanism. Those changes in the chosen transmitter entity cannot be sensed in existing EHs so broadcasted information can be only decoded at target IRs. At this point, the procedures handled by IR can be referred to  {information harvesting} which includes sending request for broadcasting to WPT via request for information (RFI) signalling, carrying out channel estimation mechanism and detecting the variations in the transmitter entity from the existing power transmitting signal. Note that the EH and IR are not necessarily  different devices so an EH can act as an information receiver. The system components of IH  can be described as follows: 

\begin{itemize}
	\item \textit{WPT} senses power request/information request from portable power receiver/portable information receiver through identification and configuration phase. Accordingly, there are two modes for wireless power transceiver, which are power transfer mode and information seeding mode. During information seeding, the process of using different transmission entity  in order to send the information to the wireless information receiver is handled. 
	\item \textit{IR} contains single/multiple receive antenna(s). Through this/them, the radiated wireless power transfer signal generated by the wireless power transceiver  is received, then; the harvesting procedure starts with estimating  corresponding transmission entity  which was the source of the information.
	\item \textit{EH} consists of receive antenna chain, battery charging unit and battery. Herein, battery charging unit  is configured to establish a link between the receive antenna chain and battery, wherein the manner in which power is transferred from wireless power transceiver is controlled in accordance with the parameters and/or state information assigned by the power management unit. 
\end{itemize}
\subsection{Improving PHY security:Dynamic activation pattern and artificial noise generation}
In order to prevent passive attacks (silent eavesdropper and eavesdropping activities), dynamic transmit entity mapping \cite{Mao2020} can be applied in order to prevent decoding the index of transmitting entities over the air by existing attacker. For dynamic transmit entity index,   dynamic transmit entity activation pattern can be added  in WPT. Those updates on transmit entity index can be done periodically after certain number of information frames based on the arrangements between WPT and IR. 

Another PHY security improvement can be obtained from artificial noise generation in full-duplex information receivers. In this scheme, information receiver can use its own sources in order to mask information transmission from wireless power transceiver. To do so, when Information receiver sends its RFI signalling, it can start to generate artificial noise signal. By this way, it will be more difficult for any existing attackers to trace the transmit entity index. Since artificial noise is generated by information receiver, the information receiver can first use a physical layer self-interference cancellation method, subtracting its own generated artificial noise signal from received signal, then information harvesting phase can start. 

Now, the details of the proposed protocol in information seeding and information harvesting phases will be described in the next section.
\section{Protocol Design}
As mentioned above, WPT initially serves a module supplying power transfer to a portable EHs and can provide information transfer to portable information receiver device(s) once  information request from the IRs is received. In order to distinguish the cycles of power transfer and information transfer at WPT,  IH protocol should have pre-signaling, identification and configuration phases.

\begin{figure*}[h]
	\centering
	\includegraphics[width=0.8\textwidth]{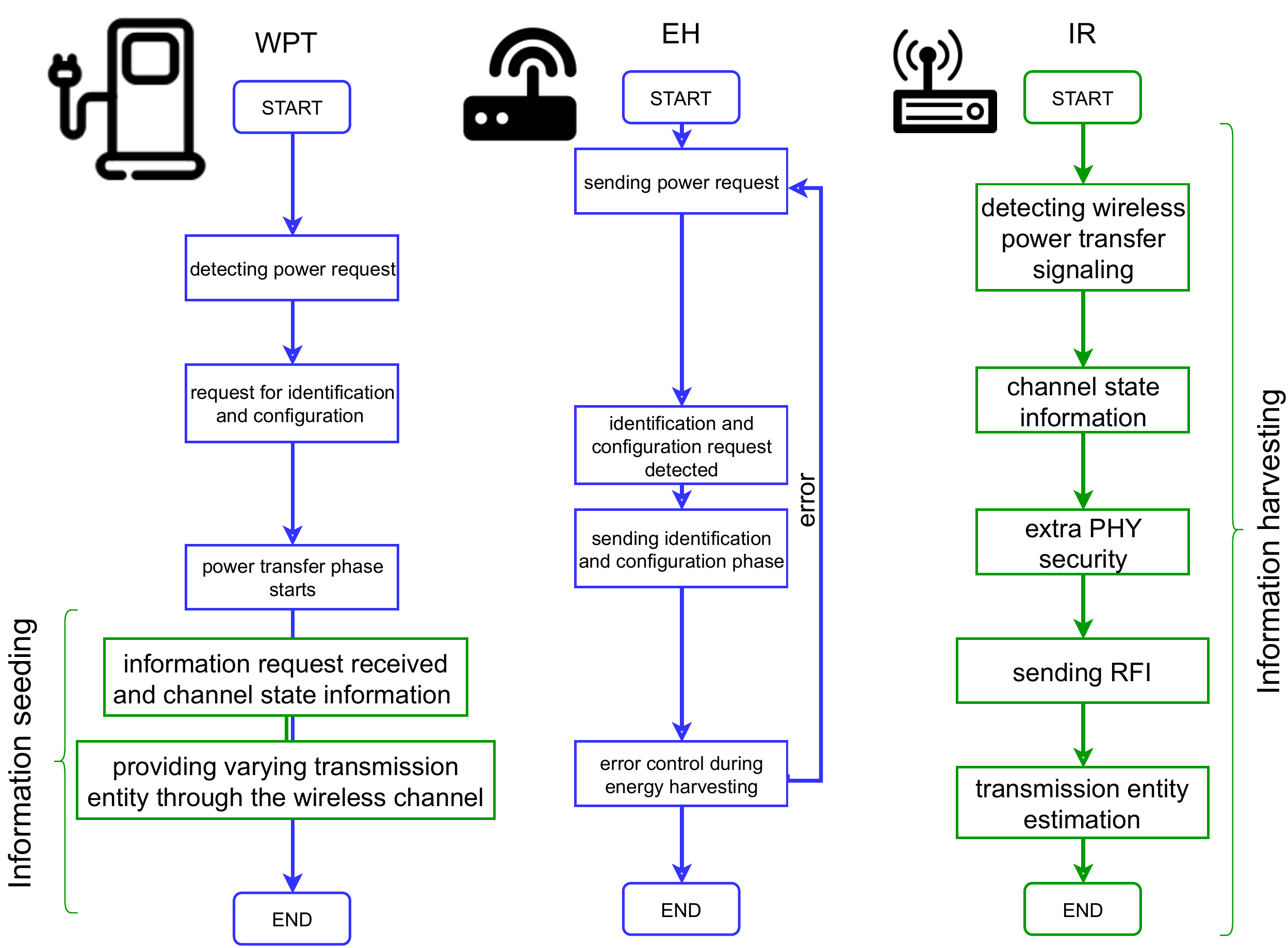}
	\caption{The detailed procedures between WPT-EH and WPT-IR during  wireless power and information transfer phases.}
	\label{fig:protocolpowertransfer}
\end{figure*}

For wireless power transfer mode when there is only EH in a service area, the \textit{ identification and configuration phases} are based on establishing wireless power transfer link between WPT and EH illustrated in Fig.~\ref{fig:protocolpowertransfer} with blue blocks. After the wireless power transfer request is initiated by EH, WPT can sense this request, and the identification of the EH and its configuration is exchanged between them. Once it is completed,  WPT unit turns into power transfer mode. Any failure in this configuration can be detected by a control error mechanism and the configuration starts from the beginning if any error is detected during power transfer period.

In case of IR enters the area, {the changes in the existing power transfer mechanism procedure is highlighted with green blocks in Fig.~\ref{fig:protocolpowertransfer}. Basically, the arrival of IR triggers two different cycles;  information seeding and information harvesting. The first one appears in WPT while the latter is executed in IR. At the beginning, information harvesting starts with sensing ongoing wireless power transfer mechanism in the service area. Once it is sensed by IR, the channel estimation process between WPT-IR links are initiated and  the parameters will be attached to RFI signal based on the channel estimation process. For instance, this process can determine the required data rate in the IR so the requirement for how frequent a chosen transmit entity is varied at WPT during information seeding can be sent through the RFI signals. Then, extra PHY security enhancement described in Section II.A can be added based on the IR preference.} 

Following those, the IR sends its information request to the  power transmitter  and this request is sensed through the receiver at the WPT. After sensing this information  request, the communications controller in the transmitter becomes  active  and the feature of  a chosen transmission entity which was determined by RF chain activation pattern is updated with respect to  information bits. Then, those changes imposed to the chosen transmission entity are estimated to extract the information embedded on wireless power signals while ongoing power transmission configuration is not affected by this procedure.

\section{Spatial modulation-based IH}

Spatial modulation (SM) technique was proposed in \cite{Renzo2011} where a subset of active transmit antennas out of multiple transmit antennas can be used in order to increase spectral efficiency. To do so, extra information bits are typically conveyed through the index of active transmit antenna(s), in addition to conventional modulated symbols. {We recall that SM was proposed as a potential NR candidate for 5G networks  \cite{3GPP} and   its derivatives,  index modulation and media-based modulation, are still popular \cite{basar2020}}.  The original version of SM only considered a single active transmit antenna over fading channels; then, generalized SM was introduced in \cite{Ishikawa2017} where a group of active transmit antennas are considered. Furthermore, there are some examples of SM proposed for non-fading/low-rank channels beyond 5G systems, i.e. mmWave channnels \cite{Liu2016}. In case of exploiting spatial modulation technique, information seeding is based on the process of using different transmit antennas at wireless power transceiver and estimating the transmit antenna in information receiver. 

\begin{figure*}[h!]
	\centering
	\includegraphics[width=\textwidth]{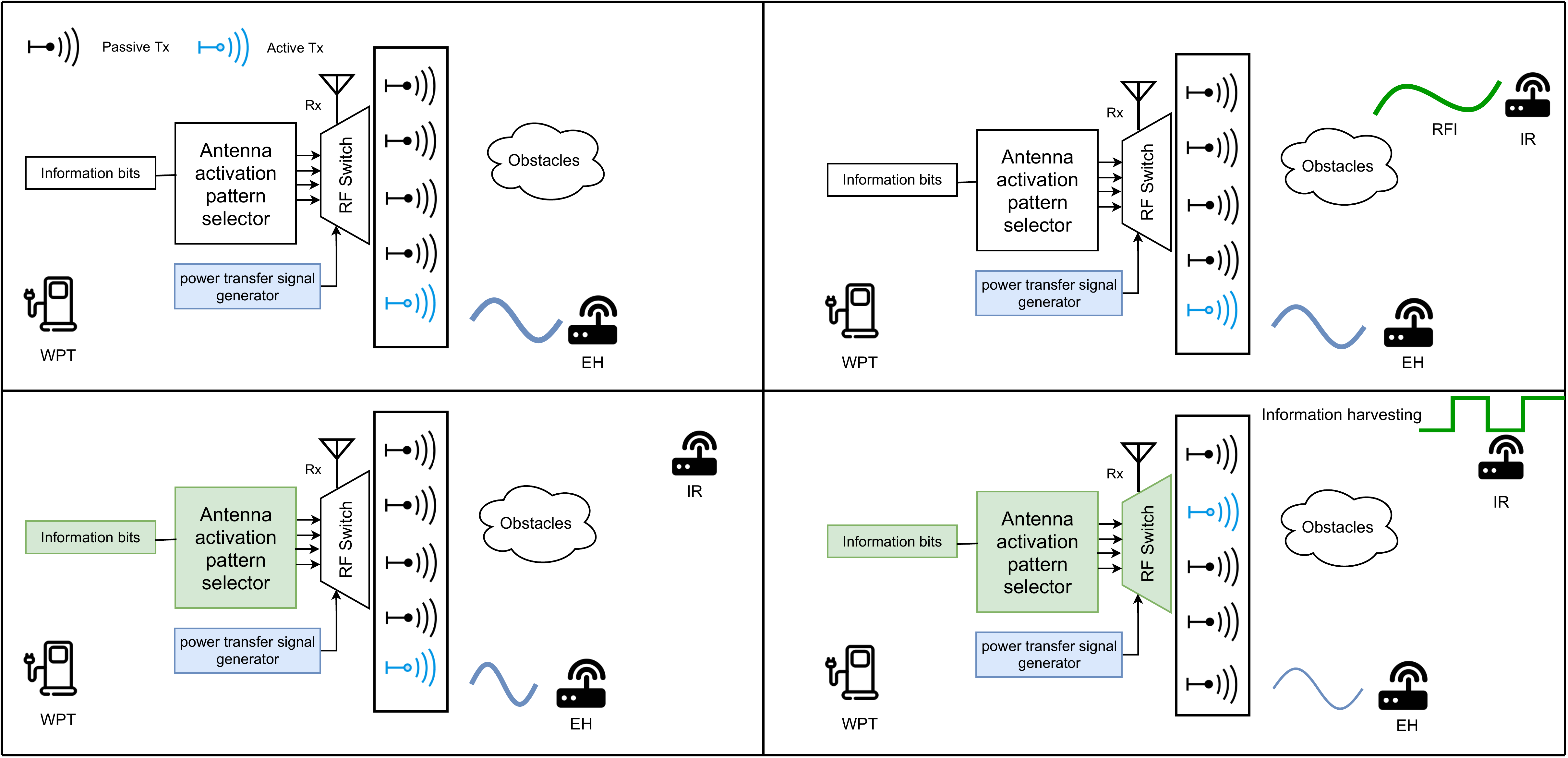}
	\caption{Spatial modulation-based Information Harvesting. (upper-left) Power transfer mode. (upper-right) IR-initiated request for information signalling. (bottom-left) Information seeding mechanism. (bottom-right) Information harvesting mechanism.}
	\label{fig:SMbasedInformationHarvesting}
\end{figure*}

In the power transfer mode, antenna activation pattern selector is inactive since there is no information receiver in the range. When the information receiver appears in the range, it transmits Reguest for Information (RFI) message. Once RFI is received in wireless power transmitting unit, antenna activation pattern selector becomes active, so wireless charging power transmitter turns into information seeding mode. {RFI signaling can also determine how fast active transmit antenna pattern varies. Recently, space-time mapping has been proposed in \cite{Irfan2020} where the period of antenna activation was extended to multiple symbol periods along with marginal increment in detection complexity.} Then, the new transmit antenna(s) can be active in the next frame based on information bits block. During the period of information seeding mode ON, information receiver aims to detect the index of active transmit antenna or indexes of group of active antennas. The overall procedure is summarized in Fig.~\ref{fig:SMbasedInformationHarvesting}.

In the case of using analog beamforming for information seeding phase, in addition to the existing power signals transmitted through EH along the selected beam pairs, the activated transmit beam indices, when correctly identified at the IR, can be used to convey additional information.  

\begin{table}[ht!]
	\caption{The simulation-related parameters}
	\centering
	\scalebox{0.95}{\begin{tabular}{|c|c|}
			\hline
			\textbf{Parameter List} & \textbf{Values} \\ \hline
			{Total transmitted power at the WPT} & 10~dB\\ \hline
			{Total active transmit antenna at the WPT} & 64\\ \hline
			The distance of information receiver & $\rm d_{\rm info}=20$ m \\ \hline
			The distance of energy harvester & $\rm d_{\rm energy}=1$ m \\ \hline
			Antenna spacing & $\lambda/2$ \\ \hline
			Operation frequency & $2$ GHz \\ \hline
			Array antenna gain & $15$ dBi \\ \hline
			Path loss model & $128.1 + 37.6\log_{10}(d)$,  d[km]\\ \hline
			Angular offset’s standard deviation & $2^{\circ}$ \\ \hline
			Log-normal shadowing’s standard deviation & $8$ dB \\ \hline
			Subcarrier bandwidth & $15$ kHz \\ \hline
			{Obstacle cover ratio} & {$0-0.9$} \\ \hline
			{Obstacle radius} & [0.3, 0.6]\\ \hline
			{Obstacle height} & [5, 25]\\ \hline
	\end{tabular}}
	\label{tab:SimulationParameters}
\end{table}

\section{Proof of Concept: Spatial Modulation based Information Harvesting}
In this section, the feasibility of Information Harvesting will be investigated through wireless power transceiver which utilizes spatial modulation mechanism. To do so, the received signal differences at information receiver and energy harvester will be investigated. 

In the wireless power transceiver,  $N_t$ transmit antennas sharing a common RF chain are employed and a group of transmit antennas are active during wireless power transmission as illustrated in Fig.~\ref{fig:SMbasedInformationHarvesting}.  In order to align with multi-antenna broadcast systems, the constant envelope power transfer signals considered in \cite{Zhang2017} are used and a set of phase shifters for each transmit antenna  is deployed. The spatial correlation between transmit antennas is modeled by the standard Kronecker correlation model which is widely used in many studies.

The characteristics of the channels for WPT-EH and WPT-IR will vary based on spatial features (near-distance and far-distance, line-of-sight and non-line-of-sight) and the WPT-IR shows more variations due to larger distance from WPT. {In order to reflect a more realistic scenario, the obstacles are modeled by  using a homogeneous Poisson point process (PPP) in the calculation of line-of-sight and non-line-of-sight probabilities for the locations of EH and IR by using obstacle coverage ratio (OCR). {Specifically, the obstacle density is determined by  the average obstacle area and total area covered by the obstacles and the obstacles are modeled as cylinders that are spatially distributed
according to a homogeneous PPP for an indoor area.} 
	
The goal is  to show the feasibility of sending information through the changes in active transmit antenna index where  energy harvesters do not notice any considerable changes in  received signals while information receiver can detect those changes in information harvesting cycle. To do so, the energy harvester is placed closer with respect to information receiver along with similar obstacle coverage ratio. The  list of the simulation parameters are given in Table~\ref{tab:SimulationParameters}.} 
\begin{figure}[t!]
	\centering
	\includegraphics[width=0.5\textwidth]{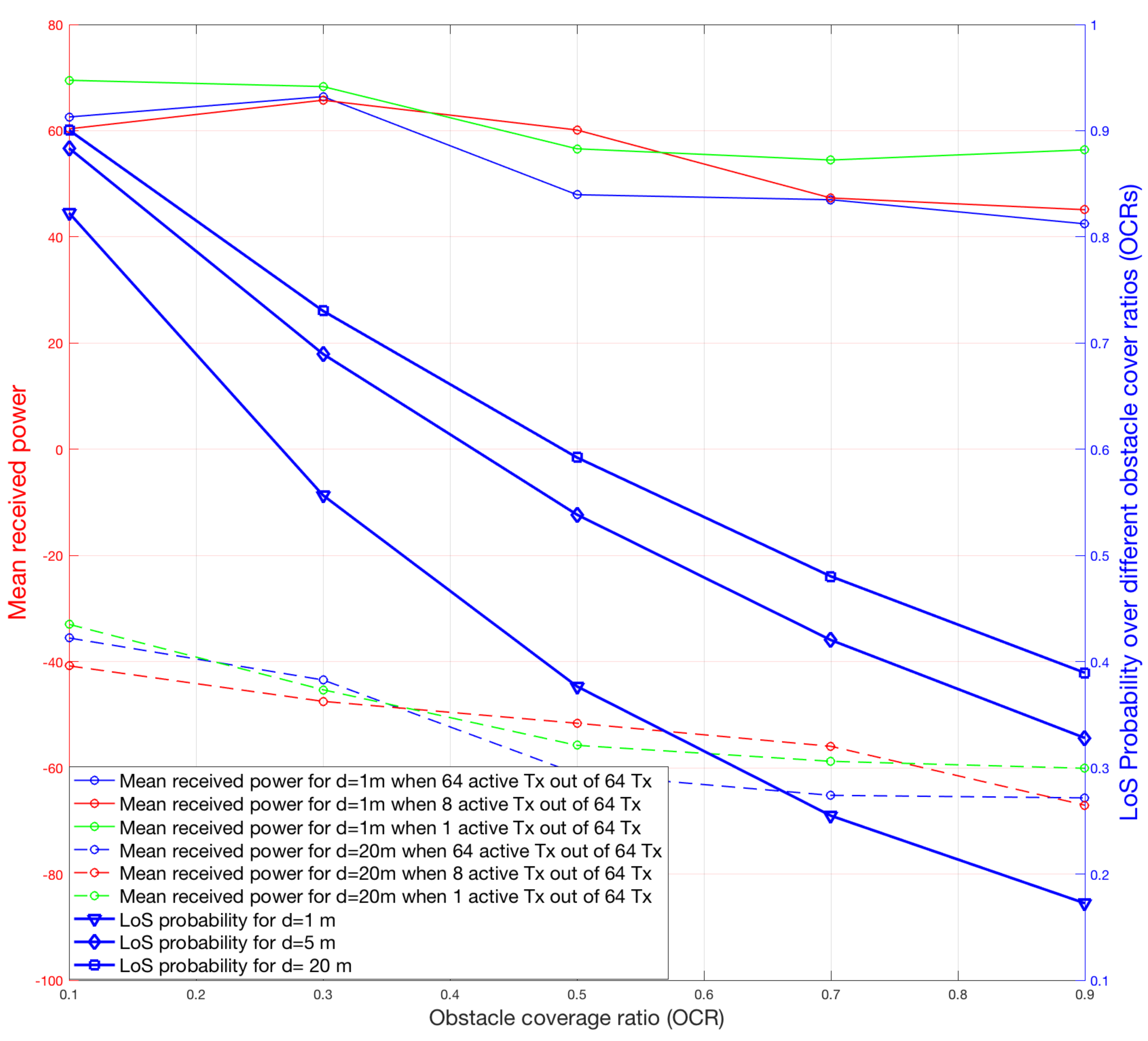}
	\caption{The received signal  variations and LoS probabilities in EH over different distances and the OCR values.}
	\label{fig:SMChannels}
\end{figure}
{The LoS probabilities with respect to different distances are plotted in Fig.~\ref{fig:SMChannels}. As seen from the curves, higher OCR values lead lower chance of getting LoS link in any distance. Now, in order to see the effect of the variations in active transmit antenna indices at wireless power transceiver, two different distances are considered for energy harvester; 1 m and 20 m. Fig.~\ref{fig:SMChannels} illustrates the mean received powers harvested in energy harvester at those distances with different number of active transmit antennas out of $64$  antennas at the power transmitter. Those average power values are obtained from $1000$ different realizations where  the positions of WPT and EH and the number of active transmit antennas are fixed.}  As seen from the figure,  changing the transmit antenna creates negligible differences in the harvested power levels and these variations result from different power transfer signal for each realization.

\begin{figure}[t!]
	\centering
	\includegraphics[width=0.48\textwidth]{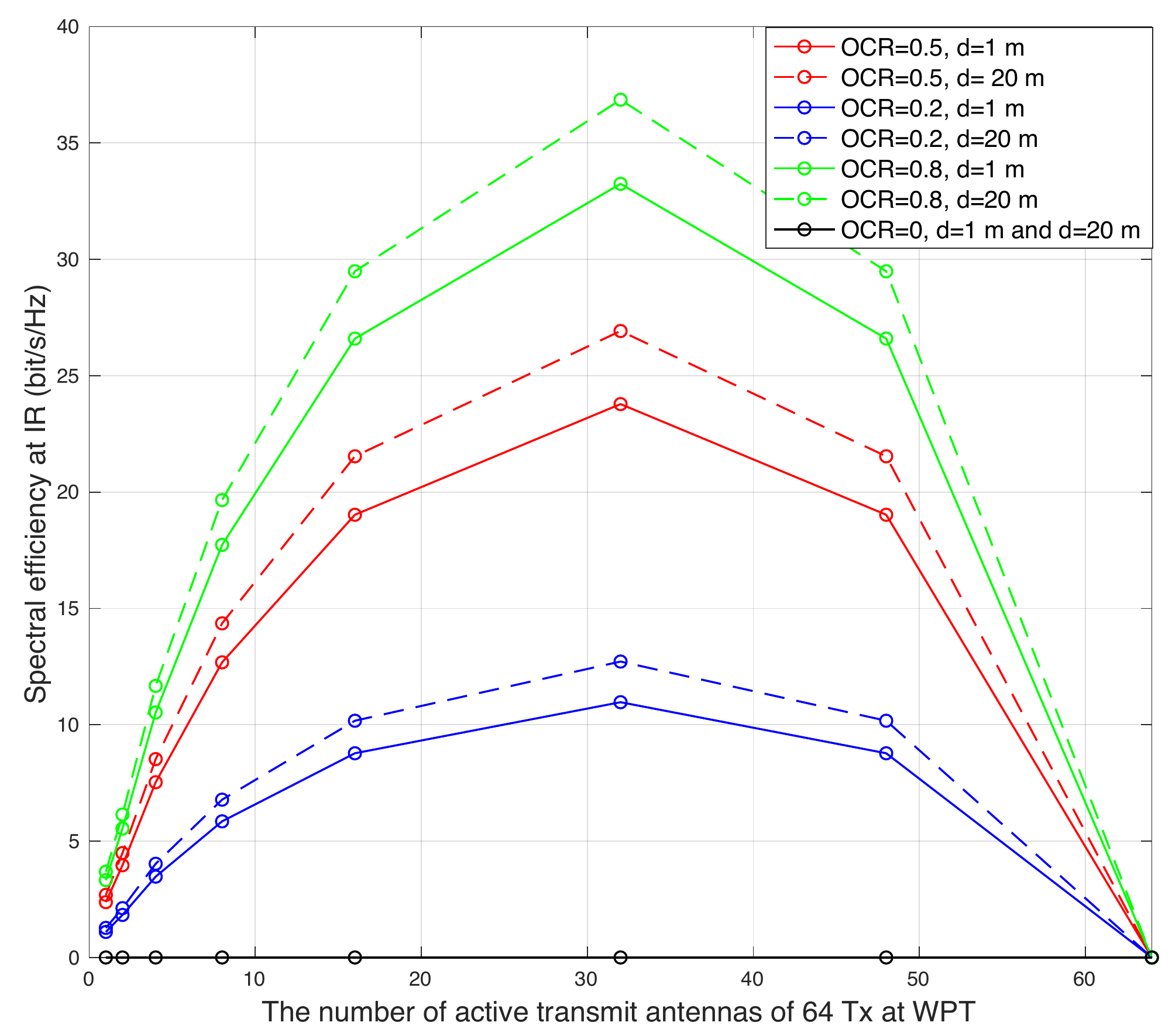}
	\caption{Upper bound for achievable spectral efficiency at IR with respect to different numbers of active transmit antennas are used at WPT.}
	\label{fig:UpperSE}
\end{figure}

The indexes of active transmit antennas during wireless power transmission period  can be detected so the extra information for information receivers can be decoded without noticing any change in energy harvesters. {The upper bound for spectral  efficiency  achieved using spatial modulation-based IH  is illustrated in Fig.~\ref{fig:UpperSE}. Therein, {different number of active transmit antennas is considered out of $64$ total transmit antenna at the WPT.} For higher number of active antennas, total combinations for information seeding is increasing until half of the antennas turn active resulting in higher spectral efficiency values at the IR. {Higher OCR and longer distance result in more variations in non-line-sight channel coefficient so it leads to higher spectral efficiencies in those cases. Interestingly, when no obstacle exists in the environment, all channels between the WPT and IR turns into roughly identical LOS channels  so spectral efficiency is zero.} Note that  Fig.~\ref{fig:UpperSE} reflects the maximum performance of IH when using  the spatial modulation-based IH. Values can be reduced in practical implementation based on different service area and device capabilities. }
\section{Conclusion}
We described a method and protocol for data communications where wireless power transfer is enhanced with a protocol to allow secure information broadcasting to devices which are in their information harvesting cycle. The key motivation is the design of information seeding and information harvesting mechanisms where energy harvesters are not able to detect the ongoing information broadcasting since harvested energy is not affected by information seeding mechanism occurring  at the wireless power transceiver. From this aspect, Information Harvesting can be utilized towards to the direction of updating IoT devices by exploiting available wireless power transfer infrastructure in the near future. In addition, Information Harvesting is a potential candidate of 3GPP Rel-18+ study items where the reduced capability devices  may be energy harvesting devices with the radio access and communication capability, thus operating on the same carrier frequencies with information harvesting devices. These devices are expected to follow a protocol allowing both energy harvesting  and information harvesting cycles.

\section{Acknowledgment}
This study has been supported by the Academy of Finland (grant number 334000).

\bibliographystyle{IEEEtran}
\bibliography{informationHarvesting}

\end{document}